\documentclass[twoside,twocolumn,9pt]{article}
\usepackage{extsizes}
\usepackage[super,sort&compress,comma]{natbib} 
\usepackage[version=3]{mhchem}
\usepackage[left=1.5cm, right=1.5cm, top=1.785cm, bottom=2.0cm]{geometry}
\usepackage{balance}
\usepackage{mathptmx,amsmath,bbold}
\usepackage{sectsty}
\usepackage{graphicx} 
\usepackage{lastpage}
\usepackage[format=plain,justification=justified,singlelinecheck=false,font={stretch=1.125,small,sf},labelfont=bf,labelsep=space]{caption}
\usepackage{float}
\usepackage{fancyhdr}
\usepackage{fnpos}
\usepackage{ upgreek }
\usepackage[english]{babel}
\addto{\captionsenglish}{%
	
}
\usepackage{array}
\usepackage{droidsans}
\usepackage{charter}
\usepackage[T1]{fontenc}
\usepackage[usenames,dvipsnames]{xcolor}
\usepackage{setspace}
\usepackage[compact]{titlesec}
\usepackage{hyperref}
\newcommand{\gj}[1]{\textcolor{black}{#1}}

\usepackage{epstopdf}

\definecolor{cream}{RGB}{222,217,201}

\begin{document}
	
	\pagestyle{fancy}
	\thispagestyle{plain}
	\fancypagestyle{plain}{
		\renewcommand{\headrulewidth}{0pt}
	}
	
	\makeFNbottom
	\makeatletter
	\renewcommand\LARGE{\@setfontsize\LARGE{15pt}{17}}
	\renewcommand\Large{\@setfontsize\Large{12pt}{14}}
	\renewcommand\large{\@setfontsize\large{10pt}{12}}
	\renewcommand\footnotesize{\@setfontsize\footnotesize{7pt}{10}}
	\makeatother
	
	\renewcommand{\thefootnote}{\fnsymbol{footnote}}
	\renewcommand\footnoterule{\vspace*{1pt}%
		\color{cream}\hrule width 3.5in height 0.4pt \color{black}\vspace*{5pt}} 
	\setcounter{secnumdepth}{5}
	
	\makeatletter 
	\renewcommand\@biblabel[1]{#1}            
	\renewcommand\@makefntext[1]%
	{\noindent\makebox[0pt][r]{\@thefnmark\,}#1}
	\makeatother 
	\renewcommand{\figurename}{\small{Fig.}~}
	\sectionfont{\sffamily\Large}
	\subsectionfont{\normalsize}
	\subsubsectionfont{\bf}
	\setstretch{1.125} 
	\setlength{\skip\footins}{0.8cm}
	\setlength{\footnotesep}{0.25cm}
	\setlength{\jot}{10pt}
	\titlespacing*{\section}{0pt}{4pt}{4pt}
	\titlespacing*{\subsection}{0pt}{15pt}{1pt}
	
	\fancyfoot{}
	\fancyfoot[LO,RE]{\vspace{-7.1pt}\includegraphics[height=9pt]{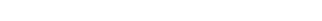}}
	\fancyfoot[CO]{\vspace{-7.1pt}\hspace{13.2cm}\includegraphics{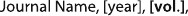}}
	\fancyfoot[CE]{\vspace{-7.2pt}\hspace{-14.2cm}\includegraphics{head_foot/RF}}
	\fancyfoot[RO]{\footnotesize{\sffamily{1--\pageref{LastPage} ~\textbar  \hspace{2pt}\thepage}}}
	\fancyfoot[LE]{\footnotesize{\sffamily{\thepage~\textbar\hspace{3.45cm} 1--\pageref{LastPage}}}}
	\fancyhead{}
	\renewcommand{\headrulewidth}{0pt} 
	\renewcommand{\footrulewidth}{0pt}
	\setlength{\arrayrulewidth}{1pt}
	\setlength{\columnsep}{6.5mm}
	\setlength\bibsep{1pt}
	
	\makeatletter 
	\newlength{\figrulesep} 
	\setlength{\figrulesep}{0.5\textfloatsep} 
	
	\newcommand{\topfigrule}{\vspace*{-1pt}%
		\noindent{\color{cream}\rule[-\figrulesep]{\columnwidth}{1.5pt}} }
	
	\newcommand{\botfigrule}{\vspace*{-2pt}%
		\noindent{\color{cream}\rule[\figrulesep]{\columnwidth}{1.5pt}} }
	
	\newcommand{\dblfigrule}{\vspace*{-1pt}%
		\noindent{\color{cream}\rule[-\figrulesep]{\textwidth}{1.5pt}} }
	
	\makeatother
	
	\twocolumn[
	\begin{@twocolumnfalse}
		{\includegraphics[height=30pt]{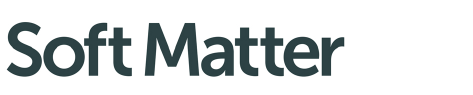}\hfill\raisebox{0pt}[0pt][0pt]{\includegraphics[height=55pt]{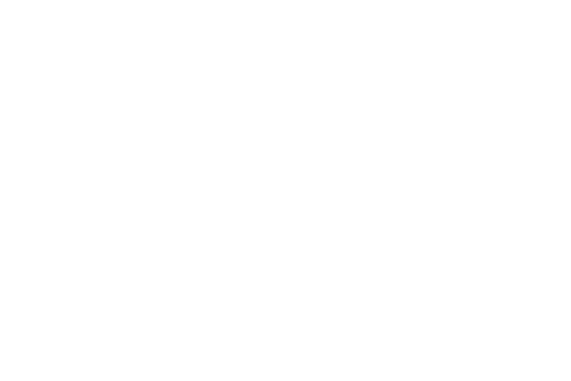}}\\[1ex]
			\includegraphics[width=18.5cm]{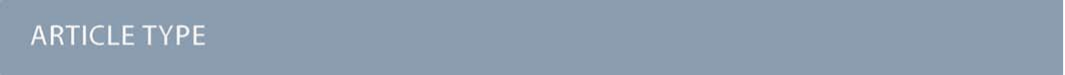}}\par
		\vspace{1em}
		\sffamily
		\begin{tabular}{m{4.5cm} p{13.5cm} }
			
			\includegraphics{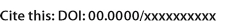} & \noindent\LARGE{\textbf{Inhomogeneous Diffusion in Confined Colloidal Suspensions}} \\
			\vspace{0.3cm} & \vspace{0.3cm} \\
			
			& \noindent\large{Gerhard Jung,$^{\ast}$\textit{$^{a\ddag}$} Alejandro Villada-Balbuena,\textit{$^{b\ddag}$} and Thomas Franosch\textit{$^{c}$}} \\
			
			\includegraphics{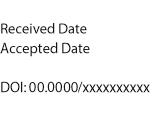} & \noindent\normalsize{		 We have performed confocal microscopy experiments and computer simulations of colloidal suspensions with moderate \gj{volume} fraction confined between two quasi-parallel, rough walls [A. Villada-Balbuena \emph{et al.}, Soft Matter, 2022, \textbf{18}, 4699-4714]. Here we investigate many facets
				of the dynamical properties of the system, such as confined and inhomogeneous diffusion, mean first-passage times and generalized incoherent scattering functions. We observe that the experiment features strong footprints of the confinement in the dynamical properties, such as inhomogeneous diffusion coefficients and non-zero off-diagonal elements in the incoherent scattering function which we can quantitatively model and analyze with computer simulations. This allows us, for example, to systematically investigate the impact of surface roughness. Our comparative study therefore advances the fundamental understanding of the impact of confinement on dynamics in fluids and colloidal suspensions.} \\
			
		\end{tabular}
		
	\end{@twocolumnfalse} \vspace{0.6cm}
	
	]
	
	\renewcommand*\rmdefault{bch}\normalfont\upshape
	\rmfamily
	\section*{}
	\vspace{-1cm}

	
	\footnotetext{\textit{$^{a}$~Laboratoire Interdisciplinaire de Physique (LIPhy), Université Grenoble Alpes, 38402 Saint-Martin-d'Hères, France; E-mail: gerhard.jung.physics@gmail.com}}
	\footnotetext{\textit{$^{b}$~Condensed Matter Physics Laboratory, Heinrich Heine University, Universit\"atsstra{\ss}e 1, 40225 D\"usseldorf, Germany; E-mail: villadab@uni-duesseldorf.de }}
		\footnotetext{\textit{$^{c}$~Institut f\"ur Theoretische Physik, Universit\"at Innsbruck, 
				Technikerstra{\ss}e 21A, 6020 Innsbruck, Austria; E-mail: thomas.franosch@uibk.ac.at }}
	
	
	\footnotetext{\ddag~These authors contributed equally to this work.}

	
	\section{Introduction}
	
	Inhomogeneous density profiles and dynamical properties in confined systems are possible since the physical confinement, for example two parallel walls, break the translational symmetry of the system. In consequence, confined fluids can behave very differently from their bulk counterparts. Important emerging phenomena are layering and confinement-induced crystallization \cite{Schmidt1997, Fortini2006, Lowen2009, Huber2015, gotzelmann1997density, nygaard2012anisotropic, Nygard2016COCIS, Nygard2016PRX, Mittal2008,D2SM00412G},
	inhomogeneous diffusion \cite{Mittal2008, Mittal2006, Granick1991, Lang2012}, and multiple-reentrant glass transitions \cite{Alcoutlabi2005, Lang2010, Mandal2014, Varnik2016, Schrack2020, Nagamanasa2015, Jung2020_A, Jung2020_B}.  Such phenomenology can be observed in a variety of different systems including atomic \cite{10.1063/1.476114}, molecular \cite{zhang1992effects} and colloidal \gj{fluids} \cite{Nugent2007}. Consequently, understanding the ramifications of confinement on the structure and dynamics of fluids  is essential for a variety of different applications \cite{10.1063/1.5057759}, including lubrication in engineering \cite{LUENGO1996328}, blood flow in biology and medicine \cite{PhysRevLett.109.108102}, as well as flow through porous media \cite{Huber_2015}. The impact of confinement on the properties of fluids has thus been investigated in the physics literature using a wide variety of different methods including experiments \cite{bechinger2002colloidal,Nugent2007,nygaard2009confinement,nygaard2012anisotropic,O_uz_2012,nygaard2013local,Edmond2012,hunter2014boundary,Nygard2016PRX,C5SM02581H,weiss2019structure,Liu2020,D2SM00419D,10.3389/fphy.2022.991540,D2SM00412G}, simulations \cite{doi:10.1063/1.457334,theo:Schmidt1996,theo:Schmidt1997,henderson1997second,alejandre1996effect,Fortini2006,doi:10.1063/1.481430,doi:10.1063/1.3623783,winkler2013computer,Mandal2014,C4SM00125G,Mandal2017,Mandal2017a,nygaard2017colloidal,Jung2020,roberts2020dynamics,jung2023mct,D4SM00339J,doi:10.1021/acs.jpcb.4c06191,doi:10.1021/acsomega.3c03624,colloids7020033} and theory \cite{antonchenko1984nature,theo:Schmidt1996,theo:Schmidt1997,doi:10.1098/rspa.2007.0115,PhysRevE.78.011602,PhysRevLett.109.240601,Lang2014a,Mandal2014,Schrack_2021,jung2023mct,jung2022extreme,10.1063/5.0156228,10.1063/5.0139116,PhysRevE.110.024601,10.1063/5.0207758}. We refer to Ref.~\cite{D2SM00412G} for a detailed introduction into this topic.
	
	Most of the above works, however, focus on structural properties of confined fluids, and thus only few is known on their dynamical properties. Notable exceptions are incoherent mode-coupling theory \cite{Jung2020_B}, computer simulations of hard-sphere fluids \cite{Mittal2006,Mittal2008,nygaard2017colloidal,jung2023mct}, and diffusion measured in experiments of confined colloidal suspensions \cite{Nugent2007,Edmond2012,hunter2014boundary,C5SM02581H}. The gist of these studies is that the inhomogeneous density profiles leave a very strong imprint on the dynamical properties of the system, in particular on the diffusion coefficient in the direction perpendicular to the confinement plane \cite{Edmond2012}. 
	
	In recent work, we have employed a combination of experimental techniques, simulations, and theoretical calculations to examine the influence of confinement on the structural properties of spherical colloids constrained between two quasi-parallel, rough walls \cite{D2SM00412G}. By modeling the short-range repulsive and medium-range screened electrostatic interactions among colloids, we achieved quantitative agreement between experimental and simulation results. This alignment enabled us to offer detailed insights into density profiles, radial distribution functions, and both anisotropic and generalized structure factors.
	
	Building on this foundation, the present manuscript adopts the same methodology to provide an in-depth analysis of essential dynamical properties, including confined mean-squared displacements, inhomogeneous diffusion, mean first-passage times, and incoherent scattering functions. We extended previous investigations of inhomogeneous diffusion in confinement by examining a broader spectrum of dynamical descriptors and by comparing experimental and numerical results on a quantitative level. The latter approach allows us to disentangle the respective contributions of structural and hydrodynamic interactions to the dynamical properties in confinement.
	
	\section{Experimental and simulation methods}
	
	In the following, we will give a brief recapitulation of the experimental and simulation methodology. For any details, we refer to Ref.~\cite{D2SM00412G} since the methodology applied in the present manuscript is the same as the one used in our previous publication.
	
	\subsection{Confocal microscopy and linking of trajectories}
	
	We perform experiments of poly(methyl metha-crylate) (PMMA) colloidal suspensions (mean diameter $\sigma_p \approx 1.85\,\upmu$m, polydispersity $\delta_p=4.8\%$) \gj{that are locked and stabilized with poly(12-hydroxy-stearic acid) (PHSA). The charges of the colloids are screened by adding  tetrabutylammonium chloride (TBAC) salt. As solvent we use a mixture of cis-decalin and cyclohexyl bromide (CHB6, purity $>98\%$, TCI), which is adjusted to be density-matched with the colloids thus avoiding, as much as possible, sedimentation of colloids\cite{D2SM00412G}. The colloids are} confined between two borosilicate cover-glass surfaces. The cover glasses are arranged to create a wedge-shaped slit with an inclination angle of less than 0.1$^{\circ}$. The surface of the glass cover is \gj{covered by melting the PMMA + PHSA + TBAC mixture onto it, thus creating a rough surfaces,} to avoid particles sticking to the glass\cite{D2SM00412G}. This creates a confinement geometry which has approximately a constant wall separation $H$ over many particle diameters, however, also allows studying different wall separations $H$ by measuring at various distant positions in the wedge. \gj{We have studied colloidal suspensions with four different \gj{volume} fractions $\varphi = N_f V_p / (L_x L_y (L + \sigma_p))$, where $N_f$ is the number of freely diffusing particles, $V_p$ is the particle volume, $L_x$ and $L_y$ are the dimensions of the box in $x$ and $y$-dimension, respectively,  and $L$ is the confinement length. Volume fractions between $\varphi=0.19$ and $\varphi=0.32$ were investigated, however,} for conciseness we focus only on the two limiting \gj{volume} fractions $\varphi=0.32$ (called `dense' in the following) and $\varphi=0.19$  (called `dilute' in the following). \gj{The given values for $\varphi$ correspond to the volume fraction of the host mixtures which have been inserted into the wedge. The inhomogeneous roughness and chemical potential due to the varying wall separation \cite{Mandal2014} will induce inhomogeneities in $\varphi$, as discussed in the results section and shown in Figs.~\ref{fig:D_dense} and \ref{fig:D_dilute}.}  As in our previous work we define the confinement length $L$ as the distance between the two liquid layers directly at the glass surface, yielding a quantity that can be easily defined for both the experiments and the simulations and is independent of wall roughness \gj{ (see Fig.~1 in Ref.~\cite{D2SM00412G} for an illustration). Approximately this definition implies that the wall separation $H \approx L + \sigma_p,$ consistent with the definition of the volume fraction above.}
	
	The samples are recorded using a confocal scanning unit. At different positions in the sample we create stacks of two-dimensional images with $512 \times 512$ pixels parallel to the glass surface which corresponds to roughly $30 \times 30$ particles. The stacks cover the whole slit from top to bottom and are recorded in vertical steps of $0.25\,\upmu$m allowing us to extract particle positions using the  interactive data language (IDL) algorithm \cite{crocker1996methods}. Up to this point the methodology is identical to the one described in detail in Ref.~\cite{D2SM00412G}.
	
	To investigate dynamical properties we link the individual measured snapshots to particle trajectories. Each scan takes about 3 to $6\,$s, depending on the wall separation, which thus defines the highest possible resolution in time for the dynamical analysis performed in this manuscript. Consequently we use $\tau= 1\,s$ as the characteristic time scale. The linking is performed using the Crocker \& Grier algorithm \cite{crocker1996methods} provided by Trackpy \cite{allan_2016_60550}. In short, the algorithm attempts to minimize the global sum of the squared displacements of particles between individual snapshots. We tried different combinations of algorithms and found that the combination of IDL for particle \gj{identification} and Trackpy for linking gave the best performance and was most efficient. After linking we have observed a drift in particle positions in all three dimensions, leading to unphysical super-diffusive mean-squared displacements. \gj{We have confirmed that this drift is caused by the motion of the probe relative to the microscope during the three hour measurements, by ensuring that the melted particles have the same drift as the freely diffusing particles.} To correct for this drift we have therefore subsequently removed the center of mass movement of the system, \gj{which proved to be more reliable than subtracting the motion of the melted wall particles which were not perfectly detected in each snapshot. We have not considered experimental trajectories beyond $t>3000\,\tau$ since imperfect linking implies that very few trajectories reach times $t > 3000\,\tau,$ leading to significant statistical errors. }
	
	\subsection{Computer simulations}
	
	We have modeled the experimental system using molecular dynamics (MD) computer simulations. We use the melted and experimentally-measured colloidal particles on the glass surfaces to create an artificial rough surface formed by frozen particles. \gj{Between these two surfaces we use the first experimentally measured snapshot to create a colloidal suspension, interacting via short-range repulsive,
	\begin{equation}\label{key}
	U(r_{ij}) = \epsilon_{\rm LJ} \left( \frac{\sigma_{\text{p},ij}}{r_{ij}} \right)^{96},
	\end{equation}
	 and medium-range electrostatic forces,
	 \begin{equation}\label{key}
	 U_Y(r_{ij}) = \frac{\epsilon_Y}{r_{ij}/\sigma_{\text{p},ij}} e^{- \kappa (r_{ij} - \sigma_{\text{p},ij} )}.
	 \end{equation}
	 Here, we have introduced the Lennard-Jones energy scale $\epsilon_{\rm LJ},$ polydisperse particle diameters $\sigma_{\text{p},ij} = (\sigma_{\text{p},i}, + \sigma_{\text{p},j})/2$ with $\sum_i \sigma_{\text{p},i} = N_f \sigma_p $ and $\text{Var}(\sigma_i) = 0.048.$ For the Yukawa potential we also introduce the Yukawa energy scale $\epsilon_Y$ and screening length $\kappa^{-1}.$ Each box consists of roughly $N_f=3000-8000$ particles, depending on the confinement lengths $L.$}
	These parameters describing the static interactions between colloids, and of colloids with the frozen wall particles, have been determined from static properties such as the inhomogeneous density profile and the radial distribution function in Ref.~\cite{D2SM00412G}. We use exactly the same parameters in the present manuscript.
	
	Particles are thermalized and kept at room temperature using a Langevin thermostat with damping time scale $\tau_L=0.1 \tau_m,$ where $\tau_m$ is the reduced time scale of the simulation model.  We do not include any explicit hydrodynamic interactions emerging from the coarse-grained fluid into the simulation model. In consequence, on short time scales, the simulated colloids will move ballistically instead of diffusively with a short-time diffusion coefficient $D_s$ as in the experiments. This implies that we cannot match a priori the reduced time scale $\tau_m$ of our simulation model to the time scale $\tau$ of the experiments. Instead, we have a single free parameter in the simulation model which we fix by comparing the mean-squared displacement measured from experiments and simulations, as will be detailed in the next section.  Comparing simulation and experimental results, thus allows us to draw conclusions on the importance of hydrodynamic interactions for the observed dynamical behavior.
	
	 \gj{Additionally, to analyze the effect of the rough boundary on the presented results, we introduce a second simulation model which will be referenced as FLAT. In this model, we remove the frozen and melted wall particles extracted from the experiments and replace them by a flat repulsive boundary which interacts with colloid $i$ according to a 48/24-WCA potential,
	 \begin{equation}
	 	V_i(r_z) = 4  \left[  \left( \frac{\sigma_{\text{p},i}}{r_{z,i}} \right)^{48} - 2  \left( \frac{\sigma_{\text{p},i}}{r_{z,i}} \right)^{24} \right],
 \end{equation}  
 where $r_{z,i}$ is the distance of the colloid from the wall.
To maintain approximately the same \gj{volume} fraction in the channel, we choose the position of the wall such that the layers adjacent to the wall are at the same position as in the case of rough boundaries.  } 

\gj{Each simulation was equilibrated and then run sufficiently long to reach the same time scales as in the experiments. With a discretized time step of $\Delta t = 10^{-4}$ this implied equilibrating for $N_{t,\text{eq}} = 2\cdot 10^4$ time steps and  simulating for $N_t = 10^8$ time steps. The total CPU cost of each simulation on our local cluster sums up to about 7 days.  }
	
	\subsection{Dynamical observables}
	
	We will characterize the dynamics of the confined colloidal suspension using various descriptors, each allowing us to highlight and better understand different aspects of confined dynamics. We will always compare results from confocal microscopy (shown as full lines in each figure) with the MD simulations (dotted lines).  Here, and in the following, $x_i(t)$ and $y_i(t)$ will denote the in-plane positions of particle $i$ at time $t,$ and $z_i(t)$ its lateral position, orthogonal to the walls. Any in-plane observable defined below will always be averaged over the two in-plane directions, even if not stated explicitly, and denoted by the subscript $_x.$ This is possible since the inclination angle of the wedge geometry in the experiments is very small and the particles have shown very similar behavior in $x-$ and $y-$direction after removing the drift. As detailed above, our trajectories are discretized with a time step $\Delta t.$
	
	\subsubsection{Mean-squared displacement and diffusion coefficients}
	
	The most basic quantity we will analyze is the mean-squared displacement (MSD),
	\begin{equation}
	\langle \Delta x(t)^2 \rangle = \frac{1}{ (N_t-m) N}\sum_{n=1}^{N_t-m}\sum_{i=1}^{N} \left [x_i(  t+n \Delta t^\prime )-x_i(n \Delta t^\prime )\right ]^2,
	\end{equation}
	\gj{for time $t=m \Delta t$. The MSD is thus averaged over all $N$ particles in the system and uses the whole trajectory by averaging over $N_t-m$ different starting times denoted by $n$.}  We similarly calculate the MSD in the confined lateral direction. The MSD thus denotes the ensemble- and time-averaged squared-displacement of the particle. \gj{Here and in the following, we have approximated the statistical error of the simulations and experiments by varying the step size $\Delta t^\prime = k \Delta t$ and evaluating the standard error of the mean for different $k$ (simple bootstrapping). Additionally, we have evaluated the standard error of the mean by calculating the variance over particles, which are largely independent, leading to similar results. We thus find that the relative errors are of the order of $1\%$. With this procedure we could thus validate that the fluctuations observed for different confinement length $L$ and the deviation between simulations and experiments is statistically relevant and caused by systematic differences such as different wall roughness and approximations made in the simulation model. }
	
	The diffusive behavior of the MSD is then fitted for \gj{$50\,\tau < t < 2000\,\tau$} using the linear function $2 D t$ to extract the longitudinal and lateral diffusion coefficients, $D_x$ and $D_z$, respectively. In lateral direction we need to restrict the time window to $t<500\,\tau$ because the lateral MSD reaches a long-time plateau (see Fig.~\ref{fig:MSD_dense}). The extracted diffusion coefficients $D_z$ thus only describe the intermittent dynamics, which is also slightly subdiffusive. From this analysis we find that choosing the simulation time scales as  $\tau_m \approx 2\,\tau$ (for $\varphi=0.32$) and $\tau_m \approx 2.8\,\tau$ (for $\varphi=0.19$) leads to a good overlap between the diffusion coefficients and will thus be used throughout the manuscript. \gj{In consequence, all simulation results are rescaled in time using the factors $2$ and $2.8$, respectively, and all figures show the experimental time scale $\tau.$ Additionally, it should be noted that the matching of time scales is based on the long-time diffusive regime. Therefore, the simulation time scale should not be used to approximate, for example, the unit of mass of the colloids, since the ballistic regimes of the experiments and simulations will likely have very different time scales. Since we solely focus on the long-time diffusive behavior here, this has no impact on the result shown in the following. }
	
	The above definition of the MSD is blind to any potential inhomogeneities in the system induced by the inhomogeneous density profile in lateral direction \cite{Mittal2008,Nugent2007,D2SM00412G}. We therefore similarly define the $z-$dependent in-plane MSD,
	\begin{equation}\label{eq:dif_space_x}
	\langle \Delta x(z,t)^2 \rangle = \frac{1}{(N_t-m) N_z}\sum_{n=1}^{N_t-m}\sum_{i\in N_z}^{} \left [ x_i( t+n \Delta t^\prime )-x_i(n \Delta t^\prime ) \right]^2.
	\end{equation}
	Here, $N_z$ includes all those particles $i$ which are within a tiny slab $[z-H_z/2,z+H_z/2]$ at time $n \Delta t^\prime.$ Thus we calculate the MSD only for such particles which start their diffusion process around the lateral position $z.$ We choose $H_z=0.15\sigma_p$.  We also calculate the same quantity in lateral direction,
	\begin{equation}\label{eq:dif_space_z}
	\langle \Delta z(z,t)^2 \rangle = \frac{1}{(N_t-m) N_z}\sum_{n=1}^{N_t-m}\sum_{i\in N_z}^{} \left[ z_i(  t+n \Delta t^\prime )-z_i(n \Delta t^\prime )\right]^2.
	\end{equation}
	From the $z-$dependent inhomogeneous MSD we extract a $z-$dependent diffusion coefficient, $D_{x/z}(z)$ using the same fitting procedure as detailed above. 
	
	\subsubsection{Mean first-passage time}
	
	To obtain even more detailed information on the inhomogeneities of the particle dynamics we calculate the expected time colloids require to travel a lateral distance $\Delta z.$ For this we define as $\tau_i(n,\Delta z)$ the time particle $i$ requires to travel a lateral distance $\left|z_i( n\Delta t^\prime+\tau_i  )-z_i(n \Delta t^\prime ) \right| = \Delta z,$ i.e. the first-passage time. Consequently, the mean first-passage time (MFPT) can be calculated as,
	\begin{equation}\label{eq:MFPT}
	\langle \tau(z,\Delta z) \rangle = \frac{1}{(N_t-m) N_z}\sum_{n=1}^{N_t-m}\sum_{i\in N_z}^{} \tau_i(n,\Delta z).
	\end{equation}
	Compared to the inhomogeneous diffusion coefficient  $D_z(z)$ the MFPT also has the important advantage that it is well defined and it does not rely on fitting an intermittent, slightly subdiffusive behavior. 
	
	\subsubsection{Incoherent scattering function}
	
	\gj{
		To capture spatio-temporal dynamics of a tracer particle in the slit, generalized intermediate scattering functions need to be introduced~\cite{Lang2012,Lang2014b,Jung2020_B,jung2023mct}. Here we recall how these objects are constructed and why they encode information on the dynamics of the particle. In a Cartesian coordinate system the position of the particle given by  the pair $(\vec{r},z)$  where $\vec{r}=(x,y)^T$ is a vector in the plane and $z\in [-H/2, H/2]$ is the transverse coordinate. The probability to find the particle at $(\vec{r},z)$ at lag time $t$ provided it was originally at $(\vec{r}', z')$ is given by the van-Hove self-correlation function, 
		\begin{equation}
		G^{(s)}(\vec{r},\vec{r}',z,z^\prime,t) = \frac{1}{N} \sum_{i=1}^N \langle  \delta ( \vec{r} - \vec{r}_i(t)  ) \delta ( \vec{r}' - \vec{r}_i(0)  ) \delta ( z - z_i(t)  ) \delta ( z^\prime - z_i(0)  )  \rangle.
		\end{equation}
		  Since the system is translationally invariant along the plane and invariant with respect to rotations in the plane, it depends only on the relative distance in the plane, but separately  on both transverse coordinates $z,z'$, i.e. $G^{(s)}= G^{(s)}(|\vec{r}-\vec{r}'|, z , z', t)$. Rather than displaying the van-Hove self-correlation function we base our discussion on its Fourier transform. The dependence on $\vec{r}-\vec{r}'$ is captured by an ordinary planar Fourier transform resulting in a dependence on a planar wave vector $\vec{q}$. By rotational invariance in the plane it depends on the its magnitude $q= |\vec{q}|$.  The dependence on $z, z'$ is encoded in discrete Fourier transforms with wavenumbers being integer multiples of $2\pi/H$. Correspondingly the generalized intermediate scattering function is defined as 
		\begin{align}
		S^{(s)}_{\mu\nu}(q,t) =& \int_{-H/2}^{H/2} d z \int_{-H/2}^{H/2} d z' \int d (\vec{r}-\vec{r}') G^{(s)}(|\vec{r}-\vec{r'}|, z, z', t) \nonumber \\
		& \times \exp( - {\rm i} Q_\mu z + {\rm i} Q_\nu z') e^{-{\rm i} \vec{q} \cdot (\vec{r}-\vec{r}') } ,
		\end{align}
		where $\mu,\nu \in \mathbb{Z}$ and $Q_\mu = 2\pi \mu/L$. The physical significance is now clear. The discrete indices $\mu, \nu$ resolve the transverse dynamics of the particle, while $q$ probes the lateral displacement of the particle. In particular, $S^{(s)}_{00}(q,t)$ reduces to the conventional ISF for the in-planar dynamics, and $S^{(s)}_{01}(q,t)$ is non-zero only if $G^{(s)}$ depends explicitly on $z$ and $z^\prime$ and not just on the difference $z - z^\prime.$  Using the microscopic definition of the van-Hove self-correlation function in terms of thermal averages of delta functions, an equivalent representation can be elaborated~\cite{Lang2014b} which amounts to evaluating
		\begin{align}
		S^{(s)}_{\mu\nu}(q,t) =& \frac{1}{(N_t-m) N}\sum_{n=1}^{N_t-m}\sum_{i=1}^{N} e^{- {\rm i} q \left[x_i(  t+n \Delta t^\prime )-x_i(n \Delta t^\prime )\right] }\\ \nonumber
		&\times \exp\left[-{\rm i} Q_\mu z_i( t+n \Delta t^\prime) \right] \exp\left[{\rm i} Q_\nu z_i(n \Delta t^\prime)\right].
		\end{align}
	}

	 For the case of symmetric walls, $S^{(s)}_{\mu\nu}(q,t)$ is a real valued quantity. We will therefore only report the real part of the ISF in this manuscript. We have validated that the imaginary part is more than one order of magnitude smaller for all experimental measurements, and statistically consistent to zero for the simulations, showing that the channel is nearly symmetric and the impact of gravity is small. This is already visible from the density profiles \cite{D2SM00412G}. Small deviations from the expected behavior could be created by asymmetries in the melted wall particles, gravity, or other experimental noise.
	
	It has been shown that for $t=0$ the ISF fulfills, $S^{(s)}_{\mu\nu}(q,t=0) = n_{\mu-\nu}$ and is thus $q-$independent \cite{Lang2014b}. Here, the density modes $n_\mu$ are calculated from the Fourier transform of the density profile,
	\begin{equation}\label{eq:den_mode}
	n_\mu=\int_{-H/2}^{H/2} dz\, e^{ -{\rm i}Q_\mu z} n(z).
	\end{equation}
	
	We have also evaluated the coherent scattering function \cite{Lang2012,Jung2020_A} but the statistics, in particular for the experimental measurements, are not sufficient to allow for a useful interpretation of the data. We have therefore decided not to include them into the manuscript. \gj{Similarly, we will focused on the lowest modes $\mu,\nu \leq 1$, since they are the natural modes to analyze the impact of confinement on the length scale $L.$ Additionally, higher order modes become increasingly noisy.}
	
	\section{Colloid dynamics in the dense sample ($\varphi=0.32$)}
	
	We will start our analysis with the dense sample which has a \gj{volume} fraction of around $\varphi=0.32,$ since we expect the impact of confinement to be most pronounced in denser systems. 
	
	\subsection{Mean-squared displacement and diffusion coefficients}
	
	\begin{figure}
		\includegraphics[width=	1.03\linewidth]{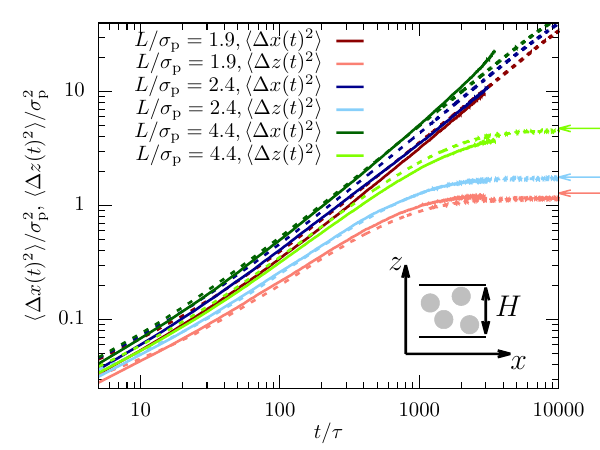}
		\caption{Mean-squared displacement in the dense sample ($\varphi=0.32$) measured using confocal microscopy (full lines) and computer simulations (dotted lines).  Shown are the in-plane ($\langle \Delta x(t)^2 \rangle$) and lateral direction  ($\langle \Delta z(t)^2 \rangle$) for channels with different confinement length $L.$ The arrows indicate the long-time limit presented in Eq.~(\ref{eq:msd_limit}). \gj{Here and in all the following figures, typical statistical errors are smaller than the line thickness.}}
		\label{fig:MSD_dense}
	\end{figure}
	
	The mean-squared displacement (MSD) highlights a fundamental difference between diffusion along the in-plane and the lateral direction \gj{for both experimental results and simulations}. While the former corresponds to free diffusion and scales as $\langle \Delta x(t \rightarrow \infty)^2 \rangle \propto t$ for long times, the latter reaches a plateau value which is directly connected to the inhomogeneous density profile,
	\begin{equation}\label{eq:msd_limit}
	\langle \Delta z(t \rightarrow \infty)^2 \rangle = \int_{-H/2}^{H/2} d z  \int_{-H/2}^{H/2} d z^\prime\, n(z) n(z^\prime) (z - z^\prime)^2,
	\end{equation}
	as accurately reproduced by  the experimental and MD data  (see Fig.~\ref{fig:MSD_dense}).
	
	In general, we observe good agreement between experiments and simulations, \gj{although small statistically relevant deviations can be observed.} This finding is non-trivial since there is only a single fit parameter to match the simulation time scale for all confinement lengths and dimensions. In particular, the good agreement implies that the long-time dynamics is \gj{only weakly influenced} by the complex hydrodynamic interactions between the colloids induced by the solvent in experiments. \gj{Instead, the dynamics is mainly dominated} by the dense packing of the colloids and their direct interactions. The very simplistic MD simulation approach using Langevin dynamics is thus sufficient to quantitatively describe the complex dynamics of dense confined colloidal suspensions.
	
	\begin{figure}
		\includegraphics[width=	1.03\linewidth]{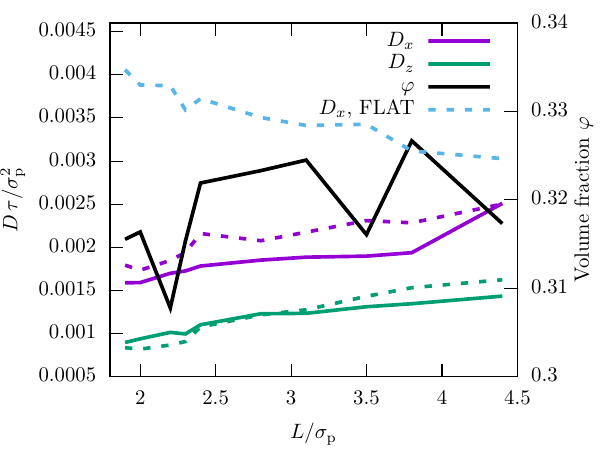}
		\caption{Diffusion coefficient $D$ along the in-plane ($x$) and the lateral ($z$) direction for different confinement lengths $L.$ The diffusion coefficient was extracted from the data in Fig.~\ref{fig:MSD_dense} using linear fits. The diffusion coefficient is compared to the \gj{volume} fraction $\varphi.$}
		\label{fig:D_dense}
	\end{figure}
	
	To investigate more systematically the impact of confinement on the dynamics, we extract the diffusion coefficient for all measured confinement lengths $L.$ We clearly observe \gj{for both techniques} that diffusion is significantly faster (about a factor of 2) along the in-plane direction compared to the lateral direction (see Fig.~\ref{fig:D_dense}). This can be explained by the fact that motion in the in-plane direction can be achieved by simply diffusing within the two-dimensional layers formed by the inhomogeneous density profile \cite{Mittal2008,D2SM00412G}. In contrast, diffusion in lateral direction requires jumping between layers. 
	
	We also clearly find in Fig.~\ref{fig:D_dense} that diffusion accelerates for larger $L,$ consistent with previous findings for hard spheres and colloids \cite{Nugent2007,Mittal2008,nygaard2017colloidal,Jung2020_B}. For soft spheres, however, the opposite effect has been observed in case of flat, smooth walls \cite{Scheidler2002}, for which strong confinement actually accelerates the dynamics. \gj{Therefore, we have also evaluated the confinement-dependent diffusion coefficient for the FLAT model, which is identical to the experimental model, just replacing the rough surface by a flat wall. Interestingly, we indeed find that this modification qualitatively changes the in-plane diffusion coefficient $D_x$ which now accelerates in strong confinement (see blue dotted curve in Fig.~\ref{fig:D_dense}). The reason for the observed behavior in the experiment is therefore very likely the rough boundary, and not the colloid interactions as in Refs.~\cite{Nugent2007,Mittal2008,nygaard2017colloidal,Jung2020_B}. }
	
	Another important subtlety observed in previous works is a non-monotonous dependence of the diffusion coefficient $D_x(L)$ on the confinement length $L$ \cite{Mittal2008,Mandal2014,nygaard2017colloidal}. This effect emerges because it is favorable for the colloids to be packed into $n$ well-defined layers (commensurate packing) rather than having many particles located between layers (incommensurate packing) as discussed in detail in Refs.~\cite{Mittal2008,Mandal2014,Jung2020_A,D2SM00412G}. We observe a similar behavior here for the MD simulations which features a \gj{very subtle} non-monotonic dependence of $D_x(L)$ oscillating on a length scale $\sigma_p,$ as expected. The effect is not very pronounced since the \gj{volume} fraction is significantly lower than in previous works and the polydispersity $\delta_p=4.8\%$ additionally weakens the effect. In contrast, the experiments do not show the same behavior and instead feature a purely monotonous dependence. The reason is, most likely, that albeit the simulations try to mimic the experiments as well as possible, including the static interactions, heterogeneous glass surfaces and polydispersity, experiments have additional sources of randomness \gj{such as polydispersity in the colloid charge}. We believe that these effects could additionally weaken the non-monotonous dependence. \gj{Furthermore, we observe that the local volume fraction depends quite non-monotonously on the position in the wedge due to locally varying wall roughness (see full black line in Fig.~\ref{fig:D_dense}). This of course leads to additional noise in the signal and thus damping of the non-monotonous behavior observed in  Refs.~\cite{Mittal2008,Mandal2014,Jung2020_A}.}
	
	\begin{figure}
		\includegraphics[width=	1.03\linewidth]{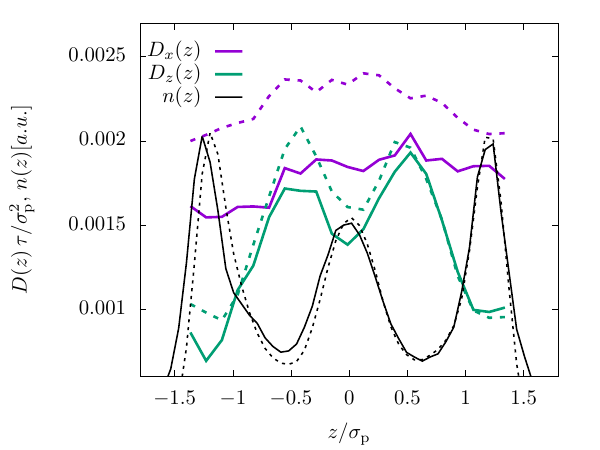}
		\caption{Position-dependent diffusion coefficient $D(z)$, as defined in Eqs.~(\ref{eq:dif_space_x}) and (\ref{eq:dif_space_z}) for the dense system ($\varphi=0.32$) and $L=2.4\sigma_p.$ Shown are the in-plane coefficient, $D_x(z),$ the lateral coefficient, $D_z(z)$, as well as the density profile $n(z).$  }
		\label{fig:Dloc_dense}
	\end{figure}
	
	Finally, we also analyze inhomogeneities in the diffusion by separating particles according to their initial lateral position. The most important finding in Fig.~\ref{fig:Dloc_dense} is the very pronounced inhomogeneity in the lateral diffusion $D_z(z)$ which shows a negative correlation with the density profile, i.e. high density implies a low diffusion coefficient \cite{Nugent2007,Mittal2008,nygaard2017colloidal}. This effect emerges because it is preferable for the colloids to be immersed inside one layer instead of being squeezed between them. Thus, if they start between two layers they quickly move towards one of the neighboring layers. Importantly, the experimental results are perfectly modeled by the MD simulations. Interestingly, this effect is completely absent in the in-plane diffusion coefficient $D_x(z)$, which shows a very weak $z$-dependence. The only visible inhomogeneity is the slightly faster diffusion in the center of the slab since the rough, frozen walls hinder the motion of the particles at the boundary.
	
	\gj{We have also analyzed the position-dependent diffusion coefficients $D(z)$ for the FLAT model, which was qualitatively identical to the results shown in Fig.~\ref{fig:Dloc_dense}. }
	
	\subsection{Mean first-passage time}
	
	\begin{figure}
		\includegraphics[width=	1.03\linewidth]{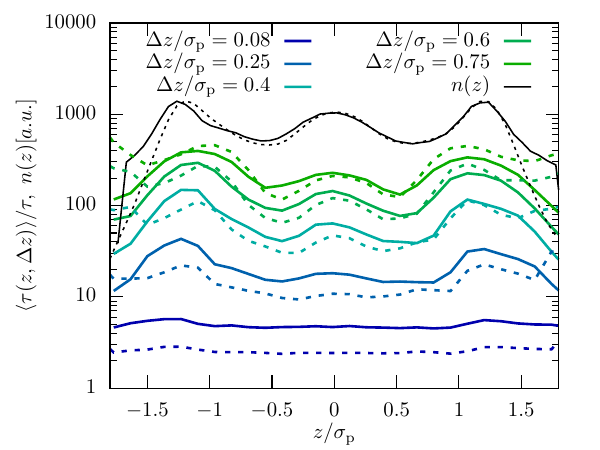}
		\caption{Mean first-passage time, $\langle \tau(z,\Delta z) \rangle $, as defined in Eq.~(\ref{eq:MFPT}) for the dense system ($\varphi=0.32$) and $L=2.4\sigma_p.$ Shown are results for different lateral distances $\Delta z,$ as well as the density profile $n(z).$ }
		\label{fig:MFPT_dense}
	\end{figure}
	
	To analyze these inhomogeneities in more detail we calculate the MFPT introduced in Eq.~(\ref{eq:MFPT}). \gj{For both techniques} we find that for very small $\Delta z \ll \sigma_p$ the MFPT is homogeneous (see Fig.~\ref{fig:MFPT_dense})\gj{, however}, there is a significant quantitative difference between experiments and MD simulations. Both observations are caused by the fact that the short-time behavior is more affected by hydrodynamic interactions with the fluid than by the inhomogeneous packing. Only when reaching $\Delta z \approx \sigma_p / 2$ the full extend of the inhomogeneities is observable and the agreement between experiments and simulations is significantly improved. This is reasonable since this length scale corresponds to the average distance a colloid has to move from between two layers to the center of a neighboring layer. In consequence, inhomogeneities are also not growing beyond $\Delta z > \sigma_p / 2,$ as seen in Fig.~\ref{fig:MFPT_dense}.
	
	\subsection{Incoherent scattering function}
	
	\begin{figure}
		\includegraphics[width=	1.03\linewidth]{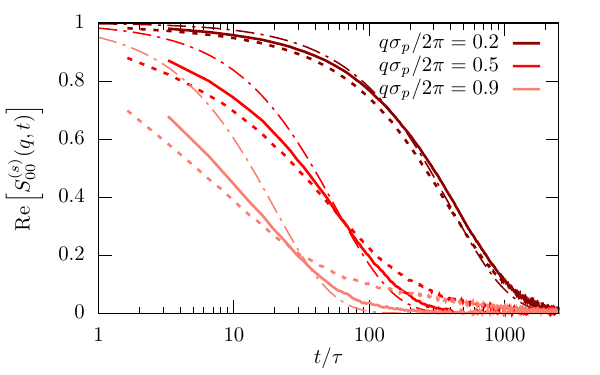}
		\caption{ Incoherent scattering function $S^{(s)}_{00}(q,t)$ for the lowest mode $\mu=\nu=0$ for $\varphi=0.32$ and $L=1.9\sigma_p$. Results for different wave numbers $q$ are obtained from confocal microscopy experiments (full lines), MD simulations (dotted lines) and the diffusive approximation Eq.~(\ref{eq:isf_dif}) using the diffusion coefficient $D$ extracted from experiments, as shown in Fig.~\ref{fig:D_dense}. }
		\label{fig:ISF_dense00}
	\end{figure}
	
\begin{figure}
		\includegraphics[width=	1.03\linewidth]{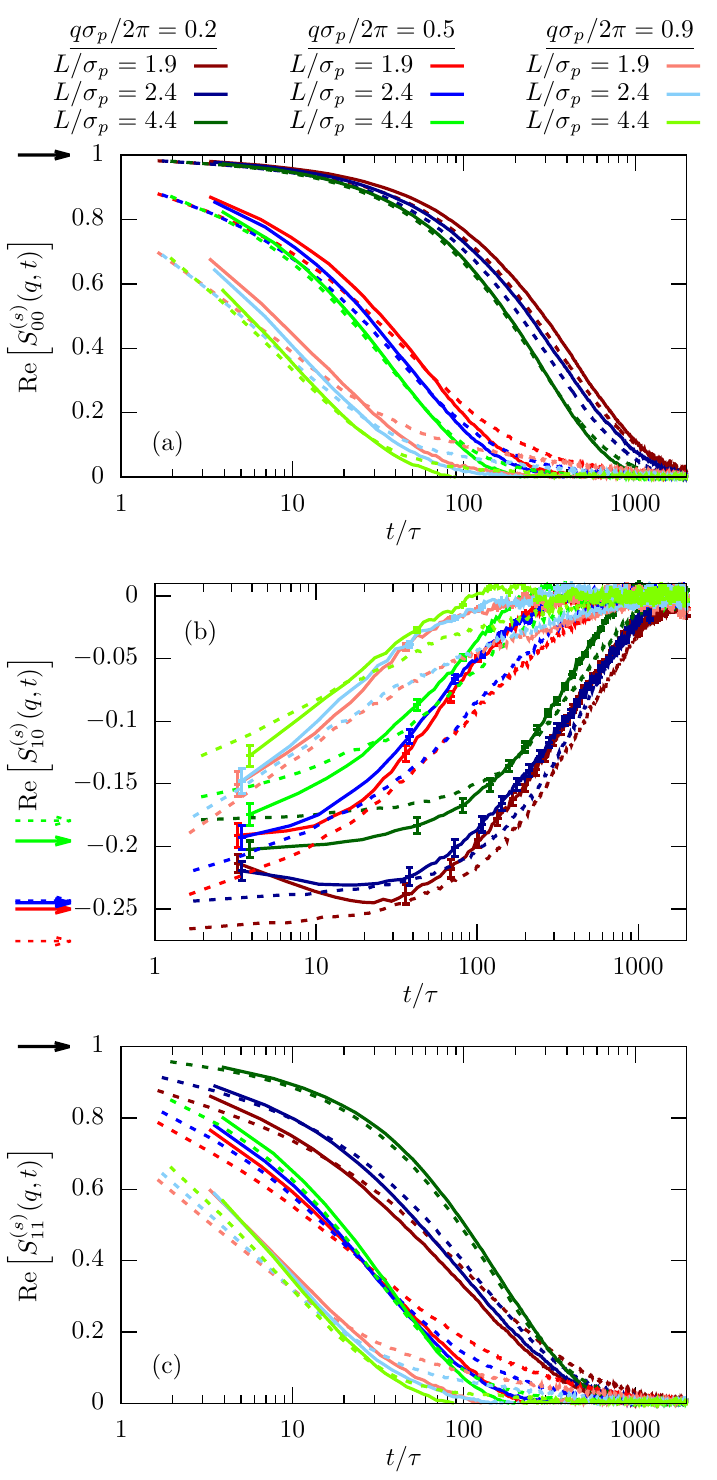}
		\caption{ Incoherent scattering function $S^{(s)}_{\mu\nu}(q,t)$ for different mode indices, $\mu=\nu=0$ (a), $\mu=0, \nu=1$ (b), $\mu=\nu=1$ (c) for $\varphi=0.32$. Results are shown for different wave numbers $q$ and various confinement lengths $L.$ The arrows indicate $n_{\mu - \nu}$, i.e. the behavior for $t \rightarrow 0$. \gj{We have included error bars in (b) for the experimental results since they are larger than the typical line width. }  }
		\label{fig:ISF_dense}
	\end{figure}

	The generalized ISF contains basically all information on the inhomogeneous diffusion process and is therefore invaluable to characterize the dynamics. The lowest mode, $\nu=\mu=0,$ integrates out any dependence on the lateral direction and thus characterizes the in-plane dynamics. For small $q$ the ISF shows a usual diffusion process corresponding to an exponential decay in time,
	\begin{equation}\label{eq:isf_dif}
	S^{(s)}_{00}(q,t)=e^{-q^2 D(L) t}
	\end{equation}
	as clearly visible in Fig.~\ref{fig:ISF_dense00} (dark red line). Experiments and simulations are in very good agreement. In contrast, for larger $q$ we observe the emergence of significant deviations between experiments and simulations. As observable in Fig.~\ref{fig:ISF_dense00}, this deviation is directly connected to the departure from the diffusive behavior shown as dashed-dotted line. \gj{The rationalization for this observation is that larger wave numbers $q$ are becoming increasingly sensitive to the behavior on small length scales and thus the interactions and dynamics on the molecular scale and fluid flow between colloids, which are both not modeled in the simulations.} In Fig.~\ref{fig:ISF_dense}a we then investigate $S^{(s)}_{00}(q,t)$ for channels with different confinement length $L.$ Consistent with the MSD we find that the diffusion process is slowed down in systems with small confinement length $L.$ 
	
	The most important feature highlighted by the ISF is the non-zero off-diagonal component $S^{(s)}_{10}(q,t).$ This behavior is only possible in systems that violate translational symmetry and feature inhomogeneous diffusion processes, as characterized above. The behavior for $t\rightarrow 0$ is well described by the density mode $n_1,$ as defined in Eq.~(\ref{eq:den_mode}) and denoted by the arrows in Fig.~\ref{fig:ISF_dense}b. We observe that the strength of the inhomogeneities increases with decreasing confinement length $L,$ as expected.
	
	Finally, we also calculate the second lowest diagonal element, $S^{(s)}_{11}(q,t).$ Its time-dependence generally follows very closely the behavior we have observed for the lowest mode $\mu=\nu=0$ (see \ref{fig:ISF_dense}c). This is expected since we have shown on quite general grounds that the dynamics of $S^{(s)}_{11}(q,t)$ couples strongly to the lowest mode \cite{Jung2020_B}.  Despite this overall similarity there are nevertheless notable differences, in particular the inverted order of the curves for the smallest wavenumber $q \sigma_p/2 \pi$ for both experiments and simulations at times $t < 500\,\tau.$ This observation highlights a notable coupling between relaxation in the in-plane and transverse directions.
	
	The agreement between simulations and experiments is generally very good for all modes of the ISF indicating that indeed all facets of colloid dynamics are very well reproduced in the simulation model. This shows how well structure and dynamics of dense colloidal suspensions can indeed be understood by simulating hard or soft, repulsive spheres, potentially with electrostatic interactions \cite{D2SM00412G,royall2024colloidal}. Even the inclusion of confinement and walls can quite easily be quantitatively incorporated into simulation models. The only real exception are the large $q$ modes which were not perfectly described in Fig.~\ref{fig:ISF_dense00}. Correctly reproducing these modes would require a more detailed simulation model, but also a molecular resolution of the confocal microscopy experiments to parameterize the model.
	
	\section{Colloid dynamics in the dilute sample ($\varphi=0.19$)}
	
	In the previous section we have found traces of the confined dynamics for a dense sample ($\varphi=0.32$) in many different dynamical observables. In the following, we will contrast these results to the dynamics in a significantly more dilute sample ($\varphi=0.19$). 
	
	\begin{figure}
		\includegraphics[width=	1.03\linewidth]{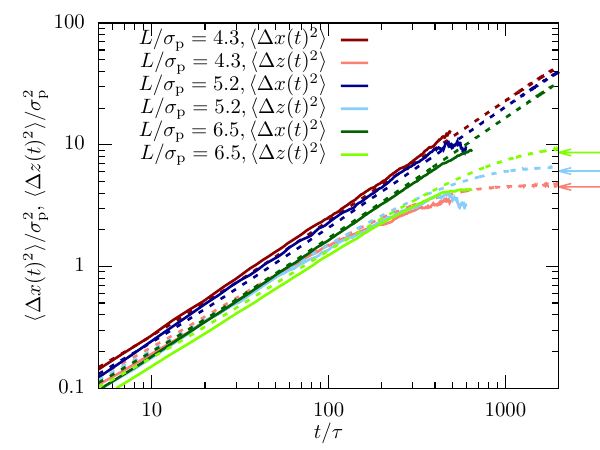}
		\caption{Mean-squared displacement in the dilute system ($\varphi=0.19$) measured using confocal microscopy (full lines) and computer simulations (dotted lines).  Shown are the in-plane ($\langle \Delta x(t)^2 \rangle$) and lateral direction  ($\langle \Delta z(t)^2 \rangle$) for channels with different confinement length $L.$ The arrows indicate the long-time limit in Eq.~(\ref{eq:msd_limit}).}
		\label{fig:MSD_dilute}
	\end{figure}
	
	The mean-squared displacement shows a similar behavior as observed for the dense sample (see Figs.~\ref{fig:MSD_dense} and \ref{fig:MSD_dilute}). In particular, we find a significantly reduced diffusion in lateral direction. The most noteworthy difference to the dense system is that the experimental trajectories show unphysical behavior for $t>600\,s.$ This is mainly because the colloid dynamics is significantly accelerated (factor 3 in the diffusion coefficient) and the scanning time increases due to the larger confinement lengths $L$ for this experiment. In consequence, the linking of the trajectories is much more difficult and leads to instabilities for longer trajectories. \gj{Additionally, particles were traveling faster and thus more quickly left the field of view of the microscope which means that there are fewer long trajectories for the dilute sample. Finally, the colloids also bleached faster since they were more exposed to the laser, making it increasingly difficult to identify particles in the later measurements.} In contrast, the simulation model does not suffer from these technical details and perfectly shows convergence towards to long-time plateau in Fig.~\ref{fig:MSD_dilute}, as predicted by Eq.~(\ref{eq:msd_limit}).
	
	\gj{We also observe in Fig.~\ref{fig:MSD_dilute} that the simulation model slightly underestimates the values for $\langle \Delta x(t)^2 \rangle$ compared to the experimental results, while it overestimates $\langle \Delta z(t)^2 \rangle$. This observation is different from the dense system in which the single time scale was sufficient to superimpose the MSD in both spatial dimensions.} We believe this is caused by the absence of hydrodynamic interactions in the simulation model which becomes more critical in dilute samples. Hydrodynamic interactions affect the in-plane and lateral dimensions differently, since the walls reflect any fluid flows and thus has a strong impact on hydrodynamics. In consequence, the assumption made in the present manuscript to just match the dynamics using a single, dimension-independent time scale becomes questionable. In other words, we can use our modeling approach to separate dynamics induced by the dense packing of colloids from the hydrodynamic interactions which are only visible in experiments.
	
	\begin{figure}
		\includegraphics[width=	1.03\linewidth]{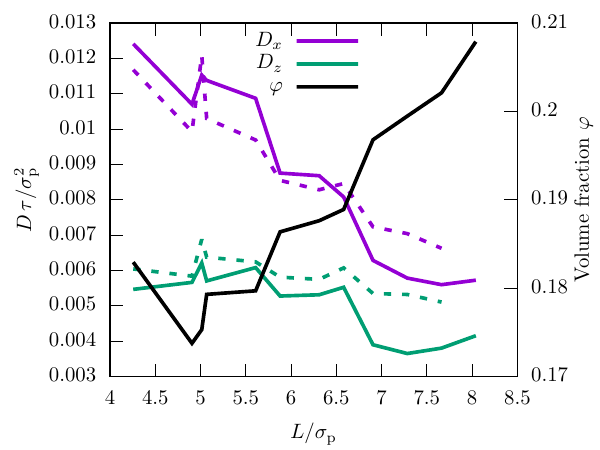}
		\caption{Diffusion coefficient $D$ along the in-plane ($x$) and the lateral ($z$) direction for different confinement lengths $L$ in the dilute sample ($\varphi=0.19$). The diffusion coefficient was extracted from the data in Fig.~\ref{fig:MSD_dilute} using linear fits. The diffusion coefficient is compared to the \gj{volume} fraction $\varphi.$}
		\label{fig:D_dilute}
	\end{figure}
	
	Extracting the diffusion coefficients $D_x$ and $D_z$ from the MSD we observe \gj{for both experiments and simulations} that diffusion becomes slower in systems with larger \gj{confinement length} $L$ (see Fig.~\ref{fig:D_dilute}). This result stands in stark contrast to the behavior discussed above for the dense system in which diffusivity was faster for larger $L.$ This surprising observation can be explained by the increase in \gj{volume} fraction $\varphi$ for larger \gj{confinement length} as shown in Fig.~\ref{fig:D_dilute}. This increase in \gj{volume} fraction $\varphi$ emerges from the wedge geometry measured in the experiments. As shown in Ref.~\cite{Mandal2014} such a wedge geometry can induce \gj{volume} fractions $\varphi(L)$ that increase with $L$ and, in polydisperse mixtures, to slower diffusion \cite{Mandal2014}. The impact of the increase in \gj{volume} fraction thus outweighs the tendency of higher diffusivity at larger $L$ observed in Fig.~\ref{fig:D_dense}.
	
	\begin{figure}
		\includegraphics[width=	1.03\linewidth]{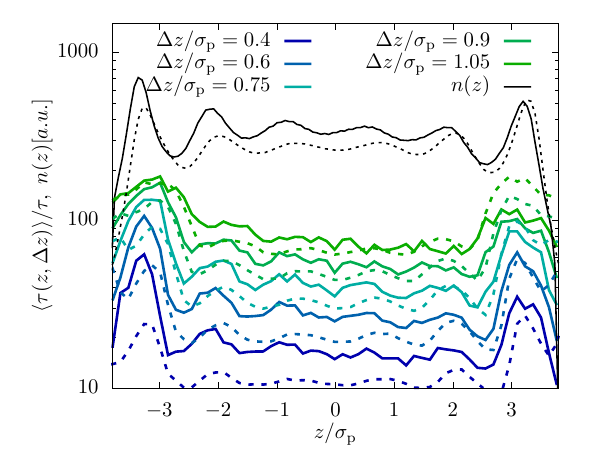}
		\caption{Mean first-passage time, $\langle \tau(z,\Delta z) \rangle $, as defined in Eq.~(\ref{eq:MFPT}) for the dilute system ($\varphi=0.19$) and $L=6.5\sigma_p.$ Shown are results for different lateral distances $\Delta z,$ as well as the density profile $n(z).$   }
		\label{fig:MFPT_dilute}
	\end{figure}
	
	There are two factors which reduce the impact of confinement on the dynamics of the dilute samples compared to the denser system: (i) the reduced \gj{volume} fraction leads to less pronounced density fluctuations which will likely also manifest itself in the dynamics, and (ii) the generally larger confinement lengths $L$ imply that, in particular in the center of the channel, the behavior is nearly bulk-like as already discussed in Ref.~\cite{D2SM00412G}. Nevertheless using the MFPT approach introduced in this manuscript we are able to visualize inhomogeneities in the dynamics, as shown in Fig.~\ref{fig:MFPT_dilute}. While in the center of the channel the MFPT is nearly flat showing that layering only plays a minor role, the impact of the pronounced boundary layer is very well visible in the dynamics. Consistent with what we have discussed for the dense system, we can thus conclude that there is a very strong correlation between the observed inhomogeneous structural properties and emerging dynamics.
	
	\begin{figure}
		\includegraphics[width=	1.03\linewidth]{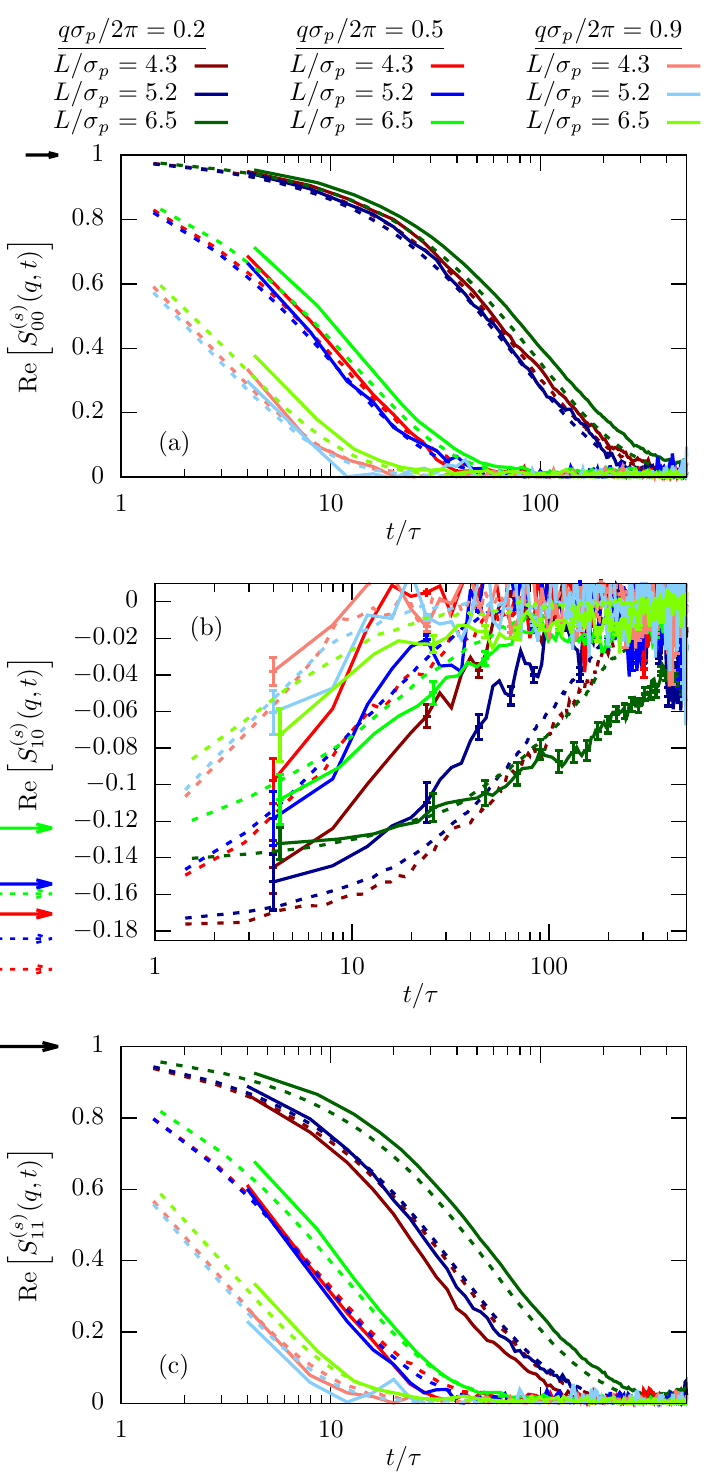}
		\caption{Incoherent scattering function $S^{(s)}_{\mu\nu}(q,t)$ for different modes $\mu=\nu=0$ (top), $\mu=0, \nu=1$ (center), $\mu=\nu=1$ (bottom) for $\varphi=0.19$. Results are shown for different wave numbers $q$ and various confinement lengths $L.$ The arrows indicate $n_{\mu - \nu}$, i.e. the behavior for $t \rightarrow 0$. \gj{We have included error bars in (b) for the experimental results since they are larger than the typical line width. } }
		\label{fig:ISF_dilute}
	\end{figure}
	
	Finally, we also investigate the incoherent scattering functions for the dilute system. Interestingly, the agreement between experiments and simulations is even better than what we observed for the dense sample (compare Figs.~\ref{fig:ISF_dense} and \ref{fig:ISF_dilute}). In particular, there is basically no discrepancy between experiments and simulations in the long-time behavior for larger $q$. We explain this by the fact that particles are less in contact in more dilute samples and thus any molecular details play a subordinate role. The simulation model thus becomes more precise in the dilute sample on the level of the static interactions. The missing hydrodynamic interactions, in return, are less important since we do not study very small $q.$
	
	It is noteworthy that we still find statistically significant deviations from zero for the off-diagonal component $S^{(s)}_{10}(q,t),$ emphasizing the importance of confinement and inhomogeneities on the colloid dynamics. However, due to the increased dilution and confinement length $L,$ the amplitude $S^{(s)}_{10}(q,t \rightarrow 0)=n_1$ is significantly smaller than in the dense system and, consequently, the signal is more noisy, in particular for the experimental results.
	
	\section{Conclusion and Outlook}
	
	We have investigated the impact of confinement on the dynamical properties of colloidal suspensions. The confinement is induced by two (nearly) parallel, rough surfaces, thus creating a channel which is just a few colloid diameters wide. In agreement with previous work, we find strong correlations between the structural properties, such as layering, and the dynamics, described by inhomogeneous diffusion coefficients and mean first-passage times. 
	
	The major contribution of the present work is that our modeling approach allows us to quantitatively compare results from confocal microscopy experiments and molecular-dynamics simulations. For most descriptors, we find \gj{good} agreement between both approaches, \gj{despite the simplicity of the simulation model. This highlights that} the dynamics of dense colloidal suspensions can be modeled using computer simulations, although both static interactions and dynamics are described only by four parameters in the coarse-grained model and no long-range hydrodynamic interactions were considered.  \gj{Larger} deviations have only been observed for the incoherent scattering function in dense suspensions for large wave numbers, and for the mean-squared displacement in dilute suspensions. We account the former to details in the short-range interactions between colloids which are not perfectly modeled by the coarse-grained simulation model, and the latter to the missing hydrodynamic interactions. \gj{Finally, replacing the rough boundary in the simulation model by a flat wall, allowed us to isolate the impact of the glass surface coating, thus showing that it qualitatively changes the confinement-dependence of the in-plane diffusion coefficient. }
	
	The goal for future experimental studies should be to go to even denser systems and investigate dynamical arrest, similar to the multi-reentrant glass transition described in Ref.~\cite{Mandal2014}. While dense systems have been studied experimentally before in Ref.~\cite{Nugent2007}, the steps  between two measured confinement lengths $L$ was too large, $\Delta L > \sigma_p$, and thus it was not possible to see any of the non-monotonous effects caused by the difference between commensurate and incommensurate packing. The challenge for such experiments will be to avoid crystallization which was observed in simulations even for high polydispersity due to fractionization induced by the walls \cite{Jung2020}.
	
	\section*{Author contributions}
	TF conceived the project. AVB performed and
	analyzed the experiments and provided input to the simulations, and GJ performed and analyzed the simulations and wrote the main draft. All authors contributed to the interpretation of the data and the writing of the manuscript.

	\section*{Data availability statement}
	
	The raw data and code used for the analysis presented in this manuscript are not available in a format that allows for public usage. Raw data and code are available from the corresponding author (GJ) upon reasonable request.
	
		\section*{Conflicts of interest}
		
		There are no conflicts to declare.
	
	\section*{Acknowledgments}
	We thank Stefan U. Egelhaaf ($\dag$2023) for initiating this project, for his supervision, for insightful discussions and just generally for his kindness. SUE always encouraged people to collaborate, resulting in fruitful projects all around the world. SUE enormous positive influence is particularly visible in the Mexican Soft Matter community due to his regular participation in Mexican conferences.
	His participation in the "25th International Conference on Science and Technology of Complex Fluids" celebrated in Puebla, Mexico, served as an inspiration for AVB to pursue a postdoctoral research stay at SUE's Experimental Soft Matter group in the Heinrich-Heine-University of Düsseldorf despite lacking an experimental background. SUE granted AVB an opportunity to work in his group, and one of the outcomes of that collaboration is this project, where AVB implemented his computational background in the analysis of Soft Matter experiments.
	This work could not exist without him. AVB acknowledges financial support provided by Conacyt (No. CVU 417675). TF acknowledges  funding by the Austrian Science Fund (FWF) 10.55776/P35673.

	\bibliography{library.bib} 
	\bibliographystyle{rsc} 
	
\end{document}